\documentclass[onecolumn, draft, 12pt]{IEEEtran}
\usepackage{amsmath,cases}
\usepackage{amsthm}
\usepackage{amssymb}
\usepackage{cite}
\usepackage{amsfonts}
\usepackage{enumerate}
\usepackage[final]{graphicx}
\usepackage{multirow}

\newcommand{\E}{{\rm E}}
\newcommand{\Cs}{\mathbb{C}}
\newcommand{\Ni}{\mathbb{N}_{i}}
\newcommand{\Ns}{\mathbb{N}}
\newcommand{\Bs}{\Cs}

\newcommand{\Cprime}{\Cs^{'}}
\newcommand{\Ss}{\Cs}
\newcommand{\G}{\mathbb{G}}
\newcommand{\R}{\mathbb{R}}

\newcommand{\Rn}{\mathbb{R}^{N}}
\newcommand{\Rnm}{\mathbb{R}^{(N-1)}}
\newcommand{\Rmn}{\mathbb{R}^{N \times (N-1) }}
\newcommand{\Rnmm}{\mathbb{R}^{(N-1) \times (N-1)}}

\newcommand{\mse}{\xi_{_{N}}}

\newcommand{\nld}{\rm NLC}

\newcommand{\Lg}{\mathcal{L}}

\newcommand{\Ltwo}{\mathcal{L}_{2}}
\newcommand{\dmp}{\mathcal{X}}

\newcommand{\onevect}{\mathbf{1}}
\newcommand{\onevectT}{\mathbf{1}^{\mathrm{T}}}

\newcommand{\La}{\mathbf{L}}
\newcommand{\D}{\mathbf{D}}
\newcommand{\B}{\mathbf{B}}
\newcommand{\A}{\mathbf{A}}
\newcommand{\I}{\mathbf{I}}

\newcommand{\Mm}{\mathbf{M}}

\newcommand{\e}{\mathbf{e}}

\newcommand{\Xcp}{\mathbf{x}_{\Cs \perp}}
\newcommand{\hXc}{\mathbf{h}_{\Cs}(\mathbf{x})}
\newcommand{\hXcp}{\mathbf{h}_{\Cs \perp}(\mathbf{x})}

\newcommand{\U}{\mathbf{U}}
\newcommand{\Sgma}{\mathbf{\Sigma}}
\newcommand{\Lmda}{\mathbf{\Lambda}}
\newcommand{\UT}{\mathbf{U}^{\mathrm{T}}}
\newcommand{\XMX}{{\mathbf{x}}^{\mathrm{T}} \mathbf{M} \mathbf{x}}
\newcommand{\XTMXt}{{\mathbf{X}(t+1)}^{\mathrm{T}} \mathbf{M} \mathbf{X}(t+1)}

\newcommand{\Czero}{\mathbf{C}}
\newcommand{\CS}{\Czero_{\rm nlc}}
\newcommand{\CSL}{\Czero_{\rm lin}}

\newcommand{\So}{\mathbf{S}}

\newcommand{\Stheta}{\So^{\theta_0}}
\newcommand{\fiN}{\mathbf{\Phi}}
\newcommand{\fiNT}{\mathbf{\Phi}^{\mathrm{T}}}

\newcommand{\xbar}{\bar{x}}

\newcommand{\x}{\mathbf{x}}
\newcommand{\X}{\mathbf{X}}
\newcommand{\xX}{\mathbf{x}}
\newcommand{\Y}{\mathbf{Y}}

\newcommand{\cval}{\ensuremath{\theta^{*}}}

\newcommand{\hX}{\mathbf{h}(\mathbf{x})}
\newcommand{\hXt}{\mathbf{h}(\mathbf{X}(t))}
\newcommand{\nt}{\mathbf{n}(t)}

\newcommand{\vX}{ V(\mathbf{x})}

\newcommand{\vecXtone}{ V(\mathbf{X}(t+1))}
\newcommand{\fitX}{ \varphi(\mathbf{x})}

\newcommand{\ntilde}{\tilde{\mathbf{n}}}
\newcommand{\ntildet}{\tilde{\mathbf{n}}(t)}
\newcommand{\xtilde}{\tilde{x}}
\newcommand{\xtildet}{\tilde{x}(t)}
\newcommand{\Xtilde}{\tilde{\mathbf{X}}}
\newcommand{\Xtildet}{\tilde{\mathbf{X}}(t)}

\newcommand{\delXt}{\delta{(\mathbf{X}(t)})}

\newcommand{\delXtildet}{\tilde{\delta}(\mathbf{X}(t)) }

\newcommand{\itN}{1 \leq i \leq N}

\newtheorem{thm}{Theorem}

\begin{document}
\title{Non-Linear Distributed Average Consensus using Bounded Transmissions}
\author{Sivaraman Dasarathan, Cihan Tepedelenlio\u{g}lu, \emph{Member, IEEE}, Mahesh Banavar, \emph{Member, IEEE} and Andreas Spanias, \emph{Fellow, IEEE}
\thanks{The authors are with the School of Electrical, Computer, and Energy Engineering, Arizona State University, Tempe, AZ 85287, USA. (Email: \{sdasarat, cihan, mbanavar, spanias\}@asu.edu). This work was supported in part by the National Science Foundation under Grant NSF FRP 1231034.}
} \maketitle

\begin{abstract}
A distributed average consensus algorithm in which every sensor transmits with bounded peak power is proposed. In the presence of communication noise, it is shown that the nodes reach consensus asymptotically to a finite random variable whose expectation is the desired sample average of the initial observations with a variance that depends on the step size of the algorithm and the variance of the communication noise. The asymptotic performance is characterized by deriving the asymptotic covariance matrix using results from stochastic approximation theory. It is shown that using bounded transmissions results in slower convergence compared to the linear consensus algorithm based on the Laplacian heuristic. Simulations corroborate our analytical findings.
\end{abstract}

\section{Introduction} \label{sec:intro_btx_consensus}
Wireless sensor networks (WSNs) without a fusion center have the advantages of robustness to node failures and they can function autonomously without a central node controlling the entire network \cite{Sankarasubramaniam2002}. In such fully distributed networks, sensors collaborate with their neighbours by repeatedly exchanging information locally to achieve a desired global objective. For example, the sensors could come to an agreement on the sample average (or on a global function) of initial measurements. This is called distributed consensus. Distributed consensus algorithms have attracted significant interest in the recent past and have found several applications in areas such as healthcare, environmental monitoring, military and home appliances (please see \cite{Boyd2003,Boyd2004,OlfatiSaber2003,OlfatiSaber2007,MinyiHuang2008,Oreshkin2008,KarMoura2009} and references therein). In this body of literature, it is often assumed that a given node can obtain exact information of the state values of its neighbours through local communications. This essentially means that the system consumes theoretically unlimited energy and bandwidth. However, practical WSNs are severely power limited and the available bandwidth is finite. Moreover, the main source of power consumption in a sensor is its transceiver \cite{PoKa05}. Therefore, there is a need for consensus algorithms which work under strict resource constraints of power and bandwidth imposed by the WSNs. 

Sensors may  adopt either a digital or analog method for transmitting their information to their neighbours. Digital methods of transmissions may be using low transmit power but require increased bandwidth especially when the number of quantization levels is high. Distributed consensus algorithms using quantized transmissions have been studied in \cite{JunFang2009,JunFang2010,AysalCoates2008,KarMoura2010,KarMoura20082}. The analog method consists of transmitting unquantized  data by appropriately pulse shaping and amplitude or phase  modulating to consume finite bandwidth. One such method is the amplify-and-forward (AF) scheme in which sensors send scaled versions of their measurements to their neighbours. However, using the AF technique is not a viable option for WSNs because it requires high transmission power when the values to be transmitted are large \cite{BanavarCM}. Moreover, the linear transmit amplifier characteristics required for AF are often very power-inefficient \cite{Bobs2012}, requiring the study of the effect of non-linear transmissions on performance. In distributed systems which employ the AF technique for transmission of the sensed data, it is often assumed that the power amplifiers used are perfectly linear over the entire range of the sensed observations. In practice, the amplifiers exhibit non-linear behaviour  when the amplitude of the sensed data is relatively high \cite{Bobs2012,Cripps2002,Cripps2006}. 

In this paper, we propose a non-linear distributed consensus ($\nld$) algorithm in which every sensor maps its state value through a bounded function before transmission to constrain the peak transmit power. Therefore the magnitude of the transmitted signal at every node in every iteration is always bounded, making it ideal for resource-constrained WSNs. In the presence of communication noise, we prove that all the sensors employing the $\nld$ algorithm reach consensus to a finite random variable whose mean is the desired sample average. We characterize the asymptotic performance by deriving the asymptotic covariance matrix using results from stochastic approximation theory. We show that using the $\nld$ algorithm results in larger asymptotic covariance compared to the linear consensus algorithm. Finally we explore the performance of the proposed algorithm employing various bounded transmission functions. Different from \cite{KarMoura2009} which also considered consensus in the presence of noisy transmissions, herein we analyse non-linear transmissions and study the asymptotic covariance matrix and its dependence on the non-linearity. Our work in this paper also studies the merits and demerits of distributed schemes involving realistic amplifier models with non-linear characteristics such as the ones discussed in \cite{Bobs2012,Cripps2002}.

The rest of this paper is organized as follows. We begin by reviewing some basics of network graph theory in Section \ref{sec:review_spectral}. In Section \ref{sec:consensus_no_noise}, we describe the system model and review the previous work on non-linear consensus. We consider the $\nld$ algorithm in the presence of noise in Section \ref{sec:consensus_with_noise}, and prove that the sensors reach consensus to a random variable. In Section  \ref{sec:simulations_nld}, we present several simulation examples to study the performance of the proposed algorithm. Concluding remarks are presented in Section \ref{Sec:Conclusions:consensus}. 

\subsection*{Notations and Conventions}\label{subsec:nld_notations}
Vectors are denoted by boldface upper-case or lower-case letters and matrices are denoted by boldface upper-case letters. $\max \lbrace a_1 , a_2\rbrace$ denotes the maximum of $a_1$ and $a_2$. ${\rm diag} [a_{1},\;  a_{2}, \; \ldots, \; a_{N}]$ denotes an $N \times N$ diagonal matrix whose diagonal elements are given by $a_{1}, a_{2}, \ldots, a_{N}$. $\E[\cdot]$ denotes the expectation operator and $\I$ denotes the identity matrix. The symbol $\| \cdot \|$ denotes the ${l}_{_{2}}$ norm for real vectors and spectral norm for symmetric matrices. For a matrix $\Mm$, $\lambda_i(\Mm)$ denotes the $i^{\rm th}$ smallest eigenvalue. The vector $\onevect$ denotes an $N \times 1$ column vector of all ones, $\onevect = [ 1 \; 1 \ldots 1]^{\rm T}$. 

\section{Review of Network Graph Theory} \label{sec:review_spectral}
In this paper, we model a sensor network as an undirected graph. In this section, we provide a brief background on network graph theory which we will use to derive our results. Consider an undirected graph $\mathbb{G}=(\mathbb{N}, \mathbb{E})$ containing a set of nodes $\mathbb{N}=\{1, \ldots, N\}$ and a set of edges $\mathbb{E}$. Nodes that communicate with each other have an edge between them. We denote the set of neighbours of node $i$ by $\mathbb{N}_{i}$, $\mathbb{N}_{i}=\{\ j|\{i,j\} \in \mathbb{E}\}$ where $\{i,j\}$ indicates an edge between the nodes $i$ and $j$ \cite{chung}. A graph is connected if there exists at least one path between every pair of nodes. We denote the number of neighbours of a node $i$ by $d_i$ and $d_{\rm max}=\max_{i} d_i$. The graph structure is described by an $N \times N$ symmetric matrix called the adjacency matrix $\A=\{a_{ij}\}$, $a_{ij}=1$ if $\{i,j\} \in \mathbb{E}$. The diagonal matrix $\D ={\rm diag} [d_{1},\;  d_{2}, \; \ldots, \; d_{N}]$ captures the degrees of all the nodes in the network. The Laplacian matrix of the graph is given by $\La=\D - \A$.

The graph Laplacian characterises a number of useful properties of the graph. The eigenvalues of $\La$ are non-negative and the number of zero eigenvalues denotes the number of distinct components of the graph. When the graph is connected, $\lambda_1 (\La) =0$, and $\lambda_i (\La) > 0 , i \geq 2$,  so that the rank of $\La$ for a connected graph is $N-1$. The vector $\onevect$ is the eigenvector of $\La$ associated with the eigenvalue $0$, i.e, $\La \onevect =\mathbf{0}$. The eigenvalue $\lambda_2 (\La)$ characterizes how densely the graph is connected and the performance of consensus algorithms depend on this eigenvalue \cite{OlfatiSaber2004}.

\section{System Model and Previous Work} \label{sec:consensus_no_noise}
\subsection{System Model}\label{subsec:nld_sys_model}
Consider a WSN with $N$ sensor nodes each with an initial measurement $x_i(0) \in \R$. Measurements made at the sensor nodes are modeled as
\begin{equation}\label{eqn:sensing_model_consensus} 
x_i (0) = \theta + n_i  \;, \hspace{0.2 in} i = 1, \ldots, N
\end{equation}
where $\theta$ is an unknown real-valued parameter and $n_i$ is the sensing noise at the $i^{\rm th}$ sensor. The sample mean of these initial measurements in \eqref{eqn:sensing_model_consensus} is given by
\begin{equation}
\label{eq:sample_avg}
\xbar= \frac{1}{N} \displaystyle\sum_{i=1}^{N} x_{i}(0) \;.
\end{equation}
Let $\xbar$ be the estimate of the parameter $\theta$ to be computed by an iterative distributed algorithm, in which each sensor communicates only with its neighbours. If the states of all the sensor nodes converge to  $\xbar$, then the network is said to have reached \emph{consensus} on the sample average. 

\subsection{Previous Work}\label{subsec:nld_prev_work}
A commonly used iterative algorithm for distributed consensus can be written as

\begin{equation}
\label{eq:LDAC}
x_{i}(t+1) = x_{i}(t) - \alpha \displaystyle\sum_{j \in \mathbb{N}_{i}} h( x_i(t) - x_{ij} (t)) \; ,
\end{equation}
where $i=1, \ldots, N$, $t=0,1,2,\ldots$, is the time index, $x_{i}(t+1)$ is the updated state value of sensor node $i$ at time $t+1$, $\Ni$ is the set of neighbours of sensor node $i$, $x_{ij} (t), j \in \mathbb{N}_{i}$ are the state values of the neighbours of sensor node $i$ at time $t$, and $\alpha$ is a constant step size. If $h(\cdot)$ is linear, then \eqref{eq:LDAC} is a linear distributed average-consensus (LDAC) algorithm \cite{Boyd2003,OlfatiSaber2004,OlfatiSaber2007}. In \cite{Boyd2003}, it is proved that if $0 < \alpha < 2 / \lambda_N(\La)$, then $x_{i}(t)$ converges to $\xbar$ exponentially and \eqref{eq:LDAC} is then called as the LDAC algorithm based on the Laplacian heuristic. If $h(\cdot)$ is non-linear then the algorithm belongs to the class of non-linear distributed average-consensus algorithms \cite{KhanKar,OlfatiSaber2003}. In \cite{OlfatiSaber2003}, the average consensus problem is solved when $h(x)$ in \eqref{eq:LDAC} is differentiable and odd. In \cite{KhanKar}, it is illustrated that when $h(x)$ in 
\eqref{eq:LDAC} is $\sin(x)$, faster convergence is possible compared to the LDAC algorithm based on the Laplacian heuristic. In all of these cases, $x_{ij} (t)$ has to be transmitted to node $i$ before it can apply the function $h(\cdot)$ to get the new updated state value. Therefore, the transmit peak power in \eqref{eq:LDAC} is determined by $x_i(t)$ and not necessarily bounded, even if $h(\cdot)$ is bounded. Moreover, there is no communication noise assumed in all the previous work on non-linear consensus.

\section{Consensus with Bounded Transmissions and Communication Noise} \label{sec:consensus_with_noise}
In this work, we propose a distributed non-linear average consensus algorithm in which every sensor maps its state value through a bounded function before transmission to constrain the transmit power. Therefore the magnitude of the transmitted signal at every node in every iteration is always bounded making it ideal for resource-constrained WSNs.

In this section, we will study the $\nld$ algorithm with communication noise when sensors exchange information. Our approach is similar to, but more general than \cite{KarMoura2009} in that we analyse non-linear transmissions. Moreover, unlike \cite{KarMoura2009} we study the asymptotic covariance matrix of the state vector and its dependence on the non-linearity. Unlike \cite{KhanKar} and \cite{OlfatiSaber2003}, we assume transmit non-linearity which allows for bounded transmissions. Moreover, we consider the presence of communication noise.

\subsection{The $\nld$ Algorithm with Communication Noise}\label{subsec:nld_with_noise}
Let each sensor map its state value at time $t$ through the function $h(x)$ before transmission, and consider the following $\nld$ algorithm with  communication noise:
\begin{equation}
\label{eq:nld_ch_noise}
x_{i}(t+1) = x_{i}(t) - \alpha(t) \displaystyle\sum_{j \in \mathbb{N}_{i} } \left [  h(x_{i}(t)) - h(x_{ij}(t)) + n_{ij}(t) \right ]\;,
\end{equation}
where $i=1, \ldots, N, t=0,1,2,\ldots$, is the time index. The value $x_{i}(t+1)$ is the state update of node $i$ at time $t+1$, $x_{ij} (t)$ is the state value of the $j^{th}$ neighbour of node $i$ at time $t$ and $\alpha(t)$ is a positive step size which will further be assumed to satisfy assumption \textbf{(A4)} in the sequel. The node $j$ transmits its information $x_{ij}(t)$ by mapping it through the function $h(x)$, node $i$ receives a noisy version of $h(x_{ij}(t))$ and $n_{ij}(t)$ is the noise associated with the reception of $h(x_{ij}(t))$. 

Note that the proposed scheme \eqref{eq:nld_ch_noise} is different from \eqref{eq:LDAC} in the following aspects. Firstly, in \eqref{eq:LDAC},  $x_{ij} (t)$ has to be transmitted which could exhibit variation over a wide range of values if $x_i(0)$ has a large dynamic range and hence \eqref{eq:LDAC} does not guarantee bounded transmission power. In contrast, in the proposed scheme the non-linearity is applied before the state value is transmitted so that the magnitude of the transmitted state value is always constrained within the maximum value of $h(x)$ irrespective of the range of $x_i(t)$ and the  realizations of noise $n_{ij}(t)$. Finally, \eqref{eq:nld_ch_noise} involves communication noise while \eqref{eq:LDAC} does not. Thus the proposed scheme is more suited to resource constrained WSNs when compared to \eqref{eq:LDAC}.

The recursion in \eqref{eq:nld_ch_noise} can be written in vector form as
\begin{equation} %
\label{eq:nld_vector_ch_noise}
\X(t+1) = \X(t) - \alpha(t) \left [ \La \hXt + \nt \right ]\;,
\end{equation}
where $\mathbf{X}(t) \in \Rn$ is the state vector at time $t$ given by $\mathbf{X}(t)=[x_{1}(t) \; x_{2}(t) \; \ldots \; x_{N}(t)]^{\rm T}$, and $\mathbf{h}: \Rn \rightarrow \Rn$ such that $\mathbf{h}(\mathbf{X}(t))=[h(x_{1}(t)) \; h(x_{2}(t)) \; \ldots \; h(x_{N}(t))]^{\rm T}$. The vector $\nt$ captures the additive noise at $N$ nodes contributed by their respective neighbours and its $i^{th}$ component is given by
\begin{equation}
\label{eq:nld_ch_noise_comp}
\mathbf{n}_{i}(t)  =  - \displaystyle\sum_{j \in \mathbb{N}_{i}} n_{ij}(t) \;, \itN \;.
\end{equation}
Our model in \eqref{eq:nld_vector_ch_noise} is more general than the linear consensus algorithm considered in \cite{KarMoura2009} which is a special case of $\hX$ when it is linear. We make the following assumptions on $h(x)$, $n_{ij}(t)$, $\alpha(t)$ and the graph:
\\
\textbf{Assumptions \;}\\
\textbf{(A1): \;} The graph $\G$ is connected so that $\lambda_2 (\La) > 0$.\\
\textbf{(A2): \;} The function $h(\cdot)$ is differentiable, and has a bounded derivative such that $0 < h^{'}(x)  \leq c$, for some $c>0$.\\ 
\textbf{(A3) Independent Noise Sequence:\;} The channel noise $\{n_{ij}(t)\}_{t\geq 0, 1\leq i,j \leq N}$ is an independent sequence across time and space. It also satisfies
\begin{align}
\label{eq:assump_A41}
\E [n_{ij}(t)]  = 0 \;, 1\leq i,j \leq N, t\geq 0\;, \; \sup_{i,j,t} \E [n_{ij}^{2}(t)]  \leq \sigma^2 < \infty.
\end{align}
From \eqref{eq:nld_ch_noise_comp} we have
\begin{align}
\label{eq:assump_A44}
\E [\nt]  = \mathbf{0} \;, \forall t\;, \; \; \mu:= \sup_{t} \E [\| \nt \|^2] \leq N d_{\rm max} \sigma^2 < \infty.
\end{align}
Note that \eqref{eq:assump_A44} is because of the fact that the number of neighbours of a given node is upper bounded by $d_{\rm max}$.\\
\textbf{(A4) Decreasing Weight Sequence:\;} The channel noise in \eqref{eq:nld_vector_ch_noise} could make the algorithm diverge. In order to control the variance growth rate of the noise we need the following conditions on the sequence $\alpha(t)$:
\begin{equation}
\label{eq:assump_A6}
\alpha(t) >0 \;, \; \displaystyle\sum_{t=0}^{\infty} \alpha(t) =  \infty \;,  \; \displaystyle\sum_{t=0}^{\infty} \alpha^{2}(t) < \infty \;.
\end{equation}

Our primary motivation for considering non-linear transmissions is to impose the realistic assumption of bounded peak per-sensor power by ensuring that 
$h(\cdot)$ is bounded. However, as seen in \textbf{(A2)} this assumption is not needed for our subsequent development as long as $h^{'}(\cdot) $ is bounded. 

We will prove convergence and asymptotic normality result of the $\nld$ algorithm in \eqref{eq:nld_vector_ch_noise}. For the sake of clarity, we now present a result on the convergence of a discrete time Markov process which will be used in establishing convergence of the $\nld$ algorithm in \eqref{eq:nld_vector_ch_noise}. 

\subsection{A Result on the Convergence of Discrete time Markov Processes}\label{subsec:conv_res_dmp}
Let $\dmp=\{\X(t)\}_{t \geq 0}$ be a discrete time vector Markov process on $\Rn$. The generating operator $\Lg$ of $\dmp$ is defined as
\begin{equation} %
\label{eq:dmp_generator}
\Lg \vX = \E \left[ \vecXtone |\X(t)=\xX \right] - \vX \;
\end{equation}
for functions $\vX, \xX \in \Rn$, provided that the conditional expectation exists. Let $\Bs \subset \Rn$ and its complement be $\Cprime = \Rn \setminus \Bs$. We now state the desired result as a simplification of Theorem 2.7.1 in \cite{Nevelson1973} (see also Theorem 1 in \cite{KarMoura2009}). 

\begin{thm} \label{nld_conv_dmp_res_thm}
Let $\dmp$ be a discrete time vector Markov process with the generator operator $\Lg$ as in \eqref{eq:dmp_generator}. If there exists a potential function $\vX : \Rn \rightarrow \R^{+}$, and $\Bs \subset \Rn$ with the following properties
\begin{align}
\label{eq:nld_conv_dmp_res_thm1}
 \vX  > 0,  \xX \in \Cprime, \;\; \vX  = 0,   \;  \xX \in \Bs \;,
\end{align}
\begin{equation} %
\label{eq:nld_conv_dmp_res_thm4}
\Lg \vX \leq - \gamma(t) \fitX + m g(t) [1+ \vX]
\end{equation}
where $m>0$, $\fitX$ is such that
\begin{align}
\label{eq:nld_conv_dmp_res_thm4a}
\fitX =0, \xX \in \Bs, \;  \fitX  > 0, \xX \in \Cprime \;,
\end{align}
and
\begin{align}
\label{eq:nld_conv_dmp_res_thm6}
\gamma(t)  > 0, g(t)  > 0, \; \displaystyle\sum_{t=0}^{\infty} \gamma(t) =  \infty, \; \displaystyle\sum_{t=0}^{\infty} g(t) < \infty \;,
\end{align}
then, the discrete time vector Markov process $\dmp=\{\X(t)\}_{t \geq 0}$ with arbitrary initial distribution converges almost surely (a.s.) to the set $\Bs$ as $t \rightarrow \infty$. That is,

\begin{equation} %
\label{eq:nld_conv_dmp_res_thm8}
{\rm Pr}\left[ \lim_{t \rightarrow \infty} \inf_{\Y \in \Bs} \; \| \X(t) - \Y \| =0 \right]=1.
\end{equation}
\end{thm}

Intuitively, Theorem \ref{nld_conv_dmp_res_thm} indicates that if the one-step prediction error of the Markov process evaluated at the potential function in \eqref{eq:dmp_generator} is bounded as in \eqref{eq:nld_conv_dmp_res_thm4} then it is possible to establish convergence of $\X(t)$.

To prove the a.s. convergence of the consensus algorithm in \eqref{eq:nld_vector_ch_noise} using Theorem \ref{nld_conv_dmp_res_thm}, we define the consensus subspace $\Cs$, the set of all vectors whose entries are of equal value as,

\begin{equation} %
\label{eq:consensus_subspace}
\Cs=\{ \xX \in \Rn | \xX = a \onevect \;, a \in \R \} \;.
\end{equation}
We are now ready to state the main result of Section \ref{sec:consensus_with_noise}.

\begin{thm} \label{nld_thm_as_conv_fixed_graph} 
Let the assumptions \textbf{(A1)}, \textbf{(A3)} and \textbf{(A4)} hold, and assume $h(x)$ is strictly increasing. Consider the $\nld$ algorithm in \eqref{eq:nld_vector_ch_noise} with the initial state vector $\X(0) \in \Rn$. Then, the state vector $\X(t)$ in \eqref{eq:nld_vector_ch_noise} approaches the consensus subspace $\Cs$ a.s., i.e.,

\begin{equation} %
\label{eq:nld_thm_as_conv_fixed_graph}
{\rm Pr}\left[ \lim_{t \rightarrow \infty} \inf_{\Y \in \Cs} \; \| \X(t) - \Y \| =0 \right]=1.
\end{equation}
\end{thm}

\begin{IEEEproof}
We will make use of Theorem \ref{nld_conv_dmp_res_thm} to prove \eqref{eq:nld_thm_as_conv_fixed_graph}. We will choose an appropriate potential function $\vX$ that is non-negative which satisfies equation \eqref{eq:nld_conv_dmp_res_thm1}. We will then prove that the generating operator $\Lg$ applied on $\vX$ as in \eqref{eq:dmp_generator} can be upper bounded as in \eqref{eq:nld_conv_dmp_res_thm4} with $\gamma(t)=\alpha(t)$, and a $\fitX$ can be found that satisfies \eqref{eq:nld_conv_dmp_res_thm4a}.

First we see that under the assumptions \textbf{(A1)}, \textbf{(A2)} and the assumption on $h(x)$, the discrete time vector process $\{\X(t)\}_{t \geq 0}$ is Markov. Since $\La$ is a positive semi-definite matrix, it has an eigenvalue decomposition (EVD) given by $\La = \U \Sgma \UT$, where  $\Sgma$ is the diagonal matrix containing the eigenvalues of $\La$ in the increasing order, and $\U$ is a unitary matrix with $\onevect$ as its first column vector which corresponds to the 0 eigenvalue. Define a positive semi-definite matrix $\Mm$ as a function of $\U$ such that $\Mm = \U \Lmda \UT$ and 
$\Lmda = {\rm diag} [0,\;  1, \; 1\;, \ldots, \; 1]$. Let $\vX=\XMX$, then the function $\vX$ is non-negative since $\Mm$ is a positive semi-definite matrix by construction. Note that $\xX \in \Cs$ is an eigenvector of $\Mm$ associated with the zero eigenvalue, therefore we have
\begin{align}
\label{eq:nld_thm_as_conv_fixed_graph2}
\vX  = 0, \xX \in \Cs \;.
\end{align}
Let $\xX = \xX_{\Ss} + \xX_{\Ss \perp}$ where $\xX_{\Ss}$ is the orthogonal projection of $\xX$ on $\Ss$. When $\xX \in \Cprime$, we have $\| \Xcp \| > 0$. Let $\xX \in \Cprime$ and $\hX$ be as defined in \eqref{eq:nld_vector_ch_noise}. Then, $\hX = \hXc + \hXcp$, where $\hXcp$ is non-zero, i.e., $\| \hXcp\| > 0$. Define $\beta := \| \hXcp\|^2 / \| \Xcp\|^2 $, then $\beta >0$, $\xX \in \Cprime$. Therefore, for any $\xX \in \Cprime$,
\begin{align}
\label{eq:nld_thm_as_conv_fixed_graph3}
\vX   = \XMX = V(\xX_{\Ss} + \xX_{\Ss \perp}) = V(\xX_{\Ss \perp}) \geq \min_{\xX_{\Ss \perp} \neq 0} \xX_{\Ss \perp}^{\rm T} \Mm \xX_{\Ss \perp} = \lambda_2 (\Mm) \| \Xcp \|^2 >0 \;,
\end{align}
where the last inequality is due to $\lambda_2 (\La) > 0$ by assumption $\textbf{(A1)}$. The equations \eqref{eq:nld_thm_as_conv_fixed_graph2} and \eqref{eq:nld_thm_as_conv_fixed_graph3} establish that the conditions in \eqref{eq:nld_conv_dmp_res_thm1} in Theorem \ref{nld_conv_dmp_res_thm} are satisfied.

Now we will prove that \eqref{eq:nld_conv_dmp_res_thm4} is satisfied as well. Towards this end, consider $\Lg \vX $ defined in \eqref{eq:dmp_generator},
\begin{align}
\label{eq:nld_conv_dmp_res_thm7a}
\Lg \vX &= \E \left[ \XTMXt |\X(t)=\xX \right] - \vX \;, \\
\nonumber
		&= \E \left[ \left( \xX^{\rm T} - \alpha(t) \left( \hX^{\rm T} \La^{\rm T} + \nt^{\rm T} \right) \right) \cdot   \left(\Mm \xX - \alpha(t) \left( \Mm \La \hX + \Mm \nt \right) \right)  \right] \\
\label{eq:nld_conv_dmp_res_thm8a}
		& \;\;\;\;\;\; - \vX \;, \\
\label{eq:nld_conv_dmp_res_thm9}
		&= - 2 \alpha(t) \left[ \xX^{\rm T} \Mm \La \hX \right] + \alpha^2(t) \left[ \hX^{\rm T} \La^{\rm T} \Mm \La \hX  + \E  \left[ \nt^{\rm T} \Mm \nt  \right] \right].
\end{align}
We get \eqref{eq:nld_conv_dmp_res_thm9} by expanding \eqref{eq:nld_conv_dmp_res_thm8a} and taking the expectations and using the fact that $\E [\nt]=\mathbf{0}$. Recall the EVDs of $\La$ and $\Mm$ from which we have

\begin{equation} %
\label{eq:nld_thm_as_conv_fixed_graph10}
\La \Mm=\Mm \La = \U \Sgma \UT \U \Lmda \UT = \U \Sgma \UT = \La  \;.
\end{equation}
Since $\lambda_2 (\Mm)=\lambda_N (\Mm)=1$, we have
\begin{equation} %
\label{eq:nld_thm_as_conv_fixed_graph11}
\E  \left[ \nt^{\rm T} \Mm \nt  \right] \leq \E  \left[ \lambda_N (\Mm) \| \nt \|^2 \right]   \leq \mu ,
\end{equation}
where the second inequality follows from \eqref{eq:assump_A44} and the fact that $\lambda_N(\Mm)=1$. Using \eqref{eq:nld_thm_as_conv_fixed_graph10} and \eqref{eq:nld_thm_as_conv_fixed_graph11} in \eqref{eq:nld_conv_dmp_res_thm9}, we get the following bound
\begin{align}
\label{eq:nld_thm_as_conv_fixed_graph12}
\Lg \vX & \leq - 2 \alpha(t) \left[ \xX^{\rm T} \La \hX \right] + \alpha^2(t) \left[ \hX^{\rm T} \La^2 \hX  + \mu \right] \;, \\
\label{eq:nld_thm_as_conv_fixed_graph13}
		& \leq - 2 \alpha(t) \left[ \xX^{\rm T} \La \hX \right] + \alpha^2(t) \left[  \lambda_N^2(\La) \beta \| \Xcp \|^2  + \mu \right] \;, \\
\label{eq:nld_thm_as_conv_fixed_graph14}
		& \leq - 2 \alpha(t) \left[ \xX^{\rm T} \La \hX \right] + \alpha^2(t) \left[ \beta \frac{\lambda_N^2(\La)}{\lambda_2(\Mm)} \XMX + \mu \right] \;, \\
\label{eq:nld_thm_as_conv_fixed_graph15}
		& \leq - 2 \alpha(t) \left[ \xX^{\rm T} \La \hX \right] + m  \alpha^2(t) \left[  1+ \beta_2 \XMX \right] \;, \\	
\label{eq:nld_thm_as_conv_fixed_graph16}
		& \leq - \alpha(t) \fitX + m  \alpha^2(t) \left[ 1+ \vX \right] \;,			
\end{align}
where $\fitX := 2 \xX^{\rm T} \La \hX$, $m := \max \{  \beta  \lambda_N^2(\La) / \lambda_2(\Mm), \mu \}$, $\beta_2 := \mu/ m$ and $\beta_2 \in (0, 1]$. In \eqref{eq:nld_thm_as_conv_fixed_graph13}, we have used the fact $\hX^{\rm T} \La^2 \hX \leq \lambda_N^2(\La) \| \hXcp \|^2$ and $\| \hXcp \|^2 = \beta \| \Xcp \|^2 $. In \eqref{eq:nld_thm_as_conv_fixed_graph14}, we have used the fact that $\XMX \geq \lambda_2(\Mm) \| \Xcp \|^2$ due to  \eqref{eq:nld_thm_as_conv_fixed_graph3}. We will now prove that $\fitX$ in \eqref{eq:nld_thm_as_conv_fixed_graph16} satisfies equation \eqref{eq:nld_conv_dmp_res_thm4a} of Theorem \ref{nld_conv_dmp_res_thm}.

Recall that $\La$ is the Laplacian matrix of the graph and that $\onevect$ is in its  null space, that is, $\La \onevect =\mathbf{0}$. Whenever $\xX \in \Cs$, i.e., $\xX = a \onevect, a \in \R$, then $\hX = b \onevect$ for some $b \in \R$. This implies $\La \mathbf{h} (a \onevect) = \La b \onevect=\mathbf{0}$. Therefore we have $\fitX=2 \xX^{\rm T} \La \hX = 0, \forall \xX \in \Cs$. 

To prove $\fitX > 0$ when $\xX \in \Cprime$, consider $\fitX$ for a connected graph with $\La$ of dimension $N \times N$,
\begin{align} 
\label{eq:nld_thm_fitX_greater_than_zero5}
\fitX & = 2 \xX^{\rm T} \La \hX \\
\label{eq:nld_thm_fitX_greater_than_zero6}
	 &  = 2 \left [ \displaystyle\sum_{j \in \Ns_{1}} (x_1-x_j) h(x_1) + \displaystyle\sum_{j \in \Ns_{2}} (x_2-x_j) h(x_2) + \ldots + \displaystyle\sum_{j \in \Ns_{N}} (x_N-x_j) h(x_N) \right ]\;, 
\end{align}
where \eqref{eq:nld_thm_fitX_greater_than_zero6} follows from the structure of the symmetric matrix $\La$ (recall $\La=\D-\A$). Note that the $i^{\rm th}$ summation in \eqref{eq:nld_thm_fitX_greater_than_zero6} corresponds to the $i^{\rm th}$ node. Now suppose that node $i$ is connected to node $j$. Then there exists a term $(x_i-x_j) h(x_i)$ in the summation corresponding to the $i^{\rm th}$ node in \eqref{eq:nld_thm_fitX_greater_than_zero6}, and a term $(x_j-x_i) h(x_j)$ in the summation corresponding to the $j^{\rm th}$ node in \eqref{eq:nld_thm_fitX_greater_than_zero6}. Both of these terms can be combined as $(x_i-x_j)(h(x_i)-h(x_j))$ and this corresponds to the edge $\{i,j\} \in \mathbb{E}$. Thus equation \eqref{eq:nld_thm_fitX_greater_than_zero6} can be written as pairwise products enumerated over all the edges in the graph as follows
\begin{align} 
\label{eq:nld_thm_fitX_greater_than_zero7}    
\fitX & =  2 \displaystyle \sum_{\{i,j\} \in \mathbb{E}} (x_i-x_j)(h(x_i)-h(x_j)) \;.
\end{align}
Since $\xX \in \Cprime$, $\fitX$ in \eqref{eq:nld_thm_fitX_greater_than_zero7} is positive due to the fact that $h(x)$ is strictly increasing so that there is at least one term in the sum which is strictly greater than zero. Letting $\gamma(t)=\alpha(t), g(t)=\alpha^2(t)$ and by assumption \textbf{(A4)}, we see that the sequence $\alpha(t)$ in \eqref{eq:nld_thm_as_conv_fixed_graph16} satisfies \eqref{eq:nld_conv_dmp_res_thm6}. Thus all the conditions of Theorem \ref{nld_conv_dmp_res_thm} are satisfied to yield \eqref{eq:nld_thm_as_conv_fixed_graph}.
\end{IEEEproof}

Theorem \ref{nld_thm_as_conv_fixed_graph} states that the sample paths of $\X(t)$ approach the consensus subspace almost surely. We note that the assumption \textbf{(A2)} is not necessary for Theorem \ref{nld_thm_as_conv_fixed_graph} to hold. Instead we assumed $h(x)$ is strictly increasing (not necessarily differentiable) to prove Theorem \ref{nld_thm_as_conv_fixed_graph}. Now, like in \cite{KarMoura2009}, we will prove the convergence of $\X(t)$ to a finite point in $\Cs$ in Theorem \ref{nld_conv_as_limiting_rv}.

\begin{thm} \label{nld_conv_as_limiting_rv}
Let the assumptions of Theorem \ref{nld_thm_as_conv_fixed_graph} hold. Consider the $\nld$ algorithm in \eqref{eq:nld_vector_ch_noise} with the initial state $\X(0) \in \Rn$. Then, there exists a finite real random variable $\cval$ such that

\begin{equation} %
\label{eq:nld_conv_as_limiting_rv}
{\rm Pr}\left[ \lim_{t \rightarrow \infty} \X(t) = \cval \onevect \right]=1.
\end{equation}

\end{thm}

\begin{IEEEproof}
Let the average of $\X(t)$ be $\xbar(t)=\onevectT \X(t) / N$. Since $\onevect \xbar(t) \in \Cs$, Theorem \ref{nld_thm_as_conv_fixed_graph} implies,

\begin{equation} %
\label{eq:nld_conv_as_limiting_rv2}
{\rm Pr}\left[ \lim_{t \rightarrow \infty} \| \X(t) -  \xbar(t) \onevect \| =0 \right]=1 \;,
\end{equation}
where \eqref{eq:nld_conv_as_limiting_rv2} follows from \eqref{eq:nld_thm_as_conv_fixed_graph} since the infimum in \eqref{eq:nld_thm_as_conv_fixed_graph} is achieved by $\Y=\xbar(t) \onevect$. Pre-multiplying \eqref{eq:nld_vector_ch_noise} by $\onevectT / N$ on both sides and noting that $\onevectT \La \hXt = \mathbf{0}$ we get, 

\begin{align}
\label{eq:nld_conv_as_limiting_rv3}
\xbar(t+1)  & =  \xbar(t) - \tilde{v}(t) \\
\label{eq:nld_conv_as_limiting_rv4}
    & =  \xbar(0) - \displaystyle\sum_{0 \leq k \leq t} \tilde{v}(k) 
\end{align}
where $\tilde{v}(t) =\alpha(t) \onevectT \nt / N$. From \textbf{(A3)}, it follows that
\begin{align}
\nonumber
\E [\tilde{v}(t)] & =0, \\
\nonumber
\label{eq:nld_conv_as_limiting_rv6}
\displaystyle\sum_{t \geq 0} \E [\tilde{v}(t)]^2 &=  \displaystyle\sum_{t \geq 0} \frac{\alpha^2(t)}{N^2} \E \| \nt \|^2 \leq  \frac{\mu}{N^2} \displaystyle\sum_{t \geq 0} \alpha^2(t) < \infty
\end{align}
which implies
\begin{equation} %
\label{eq:nld_conv_as_limiting_rv7}
\E [\xbar(t+1)]^2 \leq \xbar^2(0) + \frac{\mu}{N^2} \displaystyle\sum_{t \geq 0} \alpha^2(t)\;, \forall t \;.
\end{equation}
Equation \eqref{eq:nld_conv_as_limiting_rv7} implies that the sequence $\{\xbar(t)\}_{t \geq 0}$ is an $\Ltwo$ bounded martingale \footnote{A sequence of random variables $\{y(t)\}_{t \geq 0}$ is called as a martingale if for all $t \geq 0$, $\E\left[ |y(t)| \right] < \infty$ and $\E\left[ y(t+1) \; | \; y(1) \; y(2) \ldots y(t) \right]=y(t)$. The sequence $\{y(t)\}_{t \geq 0}$ is an $\Ltwo$ bounded martingale if $\sup_{t} \E\left[ y^2(t) \right] < \infty$ (see \cite[pp. 110]{David1991}).} and hence converges a.s. and in $\Ltwo$ to a finite random variable $\cval$ (see \cite[Theorem 2.6.1]{Nevelson1973}). Therefore the theorem follows from \eqref{eq:nld_conv_as_limiting_rv2}.
\end{IEEEproof}

It should be noted that the results in Theorems \ref{nld_thm_as_conv_fixed_graph} and \ref{nld_conv_as_limiting_rv} are similar to the results in \cite{KarMoura2009}, but we have proved it for a more general case of which \cite{KarMoura2009} is a special case when $\hX=\x$. In what follows, we present the properties of the limiting random variable $\cval$.

\subsection{Mean Square Error of $\nld$ Algorithm}\label{subsec:mse_nld}
The Theorems \ref{nld_thm_as_conv_fixed_graph} and \ref{nld_conv_as_limiting_rv} establish that the sensors reach consensus asymptotically and converge a.s. to a finite random variable  $\cval$. We can view $\cval$ as an estimate of  $\xbar$. In the following theorem we characterize the unbiasedness and means squared error (MSE) properties of $\cval$.  We define the MSE of $\cval$ as $\mse = \E [(\cval - \xbar)^2].$

\begin{thm} \label{nld_limiting_rv_mse}
Let $\cval$ be the limiting random variable as in Theorem  \ref{nld_conv_as_limiting_rv}. Then $\cval$ is unbiased, $\E [\cval] =\xbar$, and its MSE is bounded, $\mse \leq  \mu N^{-2} \displaystyle\sum_{t \geq 0} \alpha^2(t)$.
\end{thm}
The proof is obtained by following the same steps of the Lemma 5 in \cite{KarMoura2009}. 

We point out that with non-linear transmissions, we have obtained the same bound on the MSE $\mse$ as that of the linear consensus algorithm in \cite{KarMoura2009}. It should be noted that $\mu \leq N d_{\rm max} \sigma^2$ from \eqref{eq:assump_A44} which implies that $\mse \leq  d_{\rm max} N^{-1} \sum_{t \geq 0} \alpha^2(t) \sigma^2$. Therefore, if $d_{\rm max}$ is finite for a large connected network, we have $\lim_{N \rightarrow \infty} \mse =0$ and this means that $\cval$ converges to $\xbar$ as the variance of $\cval$ approaches 0. If the graph is densely connected, then $d_{\rm max}$ is relatively high which increases the worst-case MSE. On the other hand, when the graph is densely connected, $\lambda_2 (\La)$ is larger which aids in the speed of convergence to $\cval$, as quantified through the covariance matrix in Section \ref{subsec:asym_norm_nld}. 

For any connected graph with $N$ nodes, if $\sigma^2=0$ then $\lim_{t \rightarrow \infty} \X(t) = \xbar \onevect$, which means all the sensor states asymptotically converge to the desired sample average. In fact, in the absence of communication noise, under assumptions \textbf{(A1)} and \textbf{(A2)}, we believe that it is possible to prove exponential convergence of $\X(t)$ to $\xbar \onevect$ by letting $\alpha(t)=\alpha$ such that $0 < \alpha < 2 /(c \lambda_N(\La))$ and by following a similar approach as in \cite{KhanKar}.

Similar results as in Theorems \ref{nld_thm_as_conv_fixed_graph} and \ref{nld_conv_as_limiting_rv} could be easily proved under more general assumptions. For example, the graph can be randomly varying over time due to link failures. As long as the graph is connected on an average, it can be easily proved that the Theorems \ref{nld_thm_as_conv_fixed_graph} and \ref{nld_conv_as_limiting_rv} hold. The independent assumption on the noise sequence can also relaxed and the noise sequence can be allowed to depend on $\X(t)$. For detailed discussions on these assumptions and its variations, please see Section III-A in \cite{KarMoura2009}. We do not pursue these extensions herein since our focus is on studying the effect of non-linear transmissions on performance.

\subsection{Asymptotic Normality of $\nld$ Algorithm}\label{subsec:asym_norm_nld} 

The $\nld$ algorithm in \eqref{eq:nld_vector_ch_noise} belongs to the class of stochastic approximation algorithms. The convergence speed of these algorithms is an important issue from a practical perspective. There are various criteria for determining the rate of convergence. For instance, one can try to estimate $\E \left[ \| \X(t) -  \cval \onevect \|^2 \right]$ or ${\rm Pr}\left[ \| \X(t) -  \cval \onevect \| \leq \epsilon(t) \right]$ \cite{Polyak1981}. Estimating these parameters may be difficult in practice. However, it is usually possible to establish that $\sqrt{t}(\X(t) - \cval \onevect)$ is asymptotically normal with zero mean and some covariance matrix. Asymptotic normality of stochastic approximation algorithms have been established under some general conditions in \cite{Nevelson1973} and for the linear consensus algorithms in \cite{MinyiHuang2008}.

In this section, we establish the asymptotic normality of the $\nld$ algorithm in \eqref{eq:nld_vector_ch_noise}. Our approach here is similar to the one in \cite{MinyiHuang2008}. Basically, we decompose the $\nld$ algorithm in $\Rn$ into a scalar recursion and a recursion in $\Rnm$. In this section, for the sake of simplicity we assume that the noise sequence $\{ \nt, t \geq 0\}$ are i.i.d. random vectors with zero mean and finite covariance. We now formally state and prove the result as a theorem.

\begin{thm} \label{asym_norm_nld_lemma}
Let $\alpha(t)=a /t, a>0$, then the $\nld$ algorithm in \eqref{eq:nld_vector_ch_noise} becomes
\begin{equation} %
\label{eq:asym_norm_nld_lemma1}
\X(t+1) = \X(t) + \frac{a}{t} \left [ -\La \hXt + \nt \right ].
\end{equation}
Suppose that the assumptions \textbf{(A1)}, \textbf{(A2)}, \textbf{(A3)} and \textbf{(A4)} hold and that the noise sequence $\{ \nt, t \geq 0\}$ are i.i.d. across time and space with zero mean and covariance $\sigma^2_{v} \I$. Let the EVD of $\La$ be given by $\La=\U \Sgma \UT$, where $\U$ is a unitary matrix whose columns are the eigenvectors of $\La$ such that 
\begin{equation} %
\label{eq:asymp_norm_linear2} 
\U=\left [ \frac{\onevect}{\sqrt{N}} \;\; \fiN \right ], \fiN \in \Rmn \;, \; -\Sgma = \begin{bmatrix} 0  &  \mathbf{0}^{\rm T}  \\ \mathbf{0}  &  \B \\ \end{bmatrix} \;,
\end{equation} 
where $\B \in \Rnmm$ is a diagonal matrix containing the $N-1$ negative eigenvalues of $-\La$ (this means that $\B$ is a stable matrix). In addition, let $\theta_0$ be a realization of the random variable $\cval$ and $ 2 a \lambda_2(\La) h^{'}(\theta_0) >1$ so that the matrix $\left [a h^{'}(\theta_0) \B + \I /2 \right ] , \theta_0 \in \R$ is stable. Define $[\tilde{n}(t) \; \; \ntildet^{\rm T}]^{\rm T}:= N^{-1/2}  \UT \nt, \; \ntildet \in \Rnm,$ so that $\tilde{n}(t)= N^{-1}  \onevectT \nt$ and $\ntildet= N^{-1/2}  \fiNT \nt$. Let $\Czero = \E[\ntilde \ntilde^{\rm T} ]$, $\Czero \in \Rnmm$. Then, as $t \rightarrow \infty$,

\begin{equation} %
\label{eq:asym_norm_nld_lemma4}
\sqrt{t}(\X(t) - \theta_0 \onevect) \sim \mathcal{N} \left(0, N^{-1}a^2 \sigma^2_{v} \onevect \onevectT + N^{-1} \fiN \Stheta \fiNT \right) \;,
\end{equation}
where
\begin{align}
\label{eq:asym_norm_nld_lemma7}
\Stheta    & = a^2 \int\limits_{0}^{\infty} e^{\left [a h^{'}(\theta_0) \B + \frac{\I}{2} \right ] t} \; \Czero \; e^{\left [a h^{'}(\theta_0) \B + \frac{\I}{2} \right ] t} dt \;.
\end{align}
\end{thm}
\begin{IEEEproof}
Define $[\xtilde(t) \;\; \Xtilde(t)^{\rm T} ]^{\rm T}:= N^{-1/2}   \UT \X(t), \Xtilde(t) \in \Rnm$. From Theorem \ref{nld_conv_as_limiting_rv}, we have  $\X(t) \rightarrow \cval \onevect$ a.s. as $t \rightarrow \infty$ which implies that $[\xtildet \; \; \Xtildet ]^{\rm T} \rightarrow [\cval \; \; \mathbf{0} ]^{\rm T}$ a.s. as $t \rightarrow \infty$, and therefore $\Xtildet \rightarrow \mathbf{0}$ a.s. as $t \rightarrow \infty$. The error $[\X(t) - \theta_0 \onevect]$ can be written as the sum of two error components (see also Section VI in \cite{MinyiHuang2008}) as given below
\begin{align}
\label{eq:asym_norm_nld_error1}
[\X(t) - \theta_0 \onevect]  & = [\xtildet - \theta_0 ] \onevect + \frac{1}{\sqrt{N}} \fiN \Xtildet \;,\\
\label{eq:asym_norm_nld_error2}
   & = \e_1 + \e_2 \;,
\end{align}
where $\e_1=[\xtildet - \theta_0 ] \onevect$ and $\e_2= N^{-1/2} \fiN \Xtildet$. By calculating the covariance matrix between $\e_1$ and $\e_2$, it can be proved that they are asymptotically uncorrelated as $t \rightarrow \infty$, and that asymptotically $\sqrt{t} \e_1 \sim \mathcal{N} (0, N^{-1} a^2 \sigma^2_{v} \onevect \onevectT)$ (see Theorem 12 in \cite{MinyiHuang2008}). To show that $\sqrt{t} \e_2$ is asymptotically normal, it suffices to show that  $\sqrt{t} \Xtildet$ is asymptotically normal. To this end, express $h(x)$ in \eqref{eq:asym_norm_nld_lemma1} around $x=\theta_0$ using Taylor's series expansion,
\begin{equation} %
\label{eq:asym_norm_nld_proof1}
h(x) = h(\theta_0) + h^{'}(\theta_0) (x-\theta_0) + o(|x-\theta_0|)\;, {\rm as} \; x  \rightarrow \theta_0 \;.
\end{equation}
Using \eqref{eq:asym_norm_nld_proof1} in \eqref{eq:asym_norm_nld_lemma1} we get
\begin{align} %
\label{eq:asym_norm_nld_proof2}
\X(t+1) & = \X(t) + \frac{a}{t} \left [ -\La  \left(  h(\theta_0) \onevect + h^{'}(\theta_0) [\X(t) - \theta_0 \onevect ] \right)  + \delXt + \nt \right ]\;, \\
\label{eq:asym_norm_nld_proof3}
       & = \X(t) + \frac{a}{t} \left [ h^{'}(\theta_0) \left( -\La \X(t) \right)  + \delXt + \nt \right ]\;, \; {\rm as} \; t \rightarrow \infty \;,
\end{align}
where $\| \delXt\| \rightarrow 0$ as $t \rightarrow \infty$. Pre-multiplying \eqref{eq:asym_norm_nld_proof3} on both sides by $ N^{-1/2} \UT$ and using \eqref{eq:asymp_norm_linear2} we get the following recursions 
\begin{align} %
\label{eq:asym_norm_nld_proof4}
\xtilde(t+1) & = \xtildet + \frac{a}{t} \tilde{n}(t)\;, \\
\label{eq:asym_norm_nld_proof5}
\Xtilde(t+1) & = \Xtildet + \frac{a}{t} \left [ h^{'}(\theta_0) \B \Xtildet + \delXtildet + \ntildet \right ]\;, {\rm as} \; t \rightarrow \infty \;,
\end{align}
where $\delXtildet = N^{-1/2} \fiNT \delXt$. With the assumption that $\left [a h^{'}(\theta_0) \B + \I / 2 \right ] , \theta_0 \in \R$ is a stable matrix, it can be verified that all the conditions of Theorem 6.6.1 in \cite[p. 147]{Nevelson1973} are satisfied for the process $\Xtildet$ in \eqref{eq:asym_norm_nld_proof5}. Therefore, for a given $\theta_0$, the process $\sqrt{t} \Xtildet$ is asymptotically normal with zero mean and covariance matrix given by \eqref{eq:asym_norm_nld_lemma7}. Since $\sqrt{t} \e_1 \sim \mathcal{N} (0, N^{-1} a^2 \sigma^2_{v} \onevect \onevectT)$ and using \eqref{eq:asym_norm_nld_lemma7} together with the fact that $\e_1$ and $\e_2$ are asymptotically independent as $t \rightarrow \infty$, we get \eqref{eq:asym_norm_nld_lemma4} which completes the proof. 
\end{IEEEproof}

Equation \eqref{eq:asym_norm_nld_lemma4} indicates how fast the process $\X(t)$ will converge to $\theta_0 \onevect$ for a given $\theta_0$. The convergence speed clearly depends on $h^{'}(\theta_0)$. We note that if $h(x)=x$, then $h^{'}(\theta_0)=1, \forall \theta_0 \in \R$, and substituting this in \eqref{eq:asym_norm_nld_lemma7}, we get the results for the linear case as in Theorem 12 of \cite{MinyiHuang2008}.

Let the asymptotic covariance in \eqref{eq:asym_norm_nld_lemma4} be denoted by $\CS$. Since $\nt$ are i.i.d., $\Czero$ in \eqref{eq:asym_norm_nld_lemma7} becomes $\Czero=\sigma^2_{v} \I$ and thus we have $\CS= N^{-1} a^2 \sigma^2_{v} \onevect \onevectT + N^{-1} \fiN \Stheta \fiNT $ where $\Stheta$ is a diagonal matrix whose diagonal elements are given by $\Stheta_{i i} = a^2 \sigma^2_{v} / [2 a h^{'}(\theta_0) \lambda_{i+1}(\La) -1]$. A reasonable quantitative measure of largeness \cite{Polyak1981} of the asymptotic covariance matrix is  $ \| \CS \| $ which is the maximum eigenvalue of the symmetric matrix $\CS$. Further, $ \| \CS \| $ can be minimized with respect to the parameter $a$. This can be formulated as the following optimization problem,
\begin{equation} 
\label{eq:optimize_asymp_covariance}
\min_{\{a | 2 a h^{'}(\theta_0) \lambda_{2}(\La) > 1 \} } \; \max_{\{\x | \x \in \Rn,   \| \x \|^2 \leq 1 \} } \x^{\rm T} \CS \x \;,
\end{equation}
which can be solved analytically by using the KKT conditions \cite{Boyd}. The value of $a$ that optimizes \eqref{eq:optimize_asymp_covariance} is $a^{*}_{\rm nlc} = (N+1) / [ 2 N \lambda_2(\La)h^{'}(\theta_0)]$ and the corresponding optimal value of the $ \| \CS \| $is given by
\begin{equation} 
\label{eq:asymp_covariance_value}
 \| \CS^{*} \| = \left (\frac{N+1}{2 N} \right)^2 \left (\frac{\sigma^2_{v} }{\lambda^2_2(\La)} \right )  \left ( \frac{1}{h^{'}(\theta_0)} \right )^2 \;.
\end{equation}
The size of the asymptotic covariance matrix in \eqref{eq:asymp_covariance_value} is inversely proportional to the square of the smallest non-zero eigenvalue $\lambda_2(\La)$ which quantifies how densely a graph is connected. We also note that \eqref{eq:asymp_covariance_value} is directly proportional to the channel noise variance $\sigma^2_{v}$. 

Equation \eqref{eq:asymp_covariance_value} also gives some useful insights to design the transmission function $h(x)$. If we choose two functions $h_1(x)$ and $h_2(x)$ such that $h^{'}_{1}(x) > h^{'}_{2}(x), \forall x \in \R$, it is easy to see from \eqref{eq:asymp_covariance_value} that $ \| {\CS^{*}}_1 \| < \| {\CS^{*}}_2 \|, \forall \theta_0 \in \R$. This means that the convergence will be faster when $h_1(x)$ is employed in the $\nld$ algorithm \eqref{eq:nld_vector_ch_noise} than when $h_2(x)$ is employed. However, it should be noted that if $h^{'}_{1}(x) > h^{'}_{2}(x), \forall x \in \R$ and suppose $h_{1}(0) = h_{2}(0)=0$ then we have $h^{2}_{1}(x) > h^{2}_{2}(x), \forall x$ which implies that on an average the transmit power is greater when $h_1(x)$ is employed compared to $h_2(x)$. We will illustrate these findings in the simulations in Section \ref{sec:simulations_nld}. Comparing the  $\| \CS^{*} \|$ against the special case of $h(x)=c x$ yields $\| \CS^{*} \| = \| \CSL^{*} \| (c/h^{'}(\theta_0))^2$. Clearly $c/h^{'}(\theta_0) \leq 1$ and therefore if $h(x)$ is bounded, appropriately normalized by letting $c=1$, so that $0 < h^{'}(x) \leq 1$, we conclude that the best case linear algorithm outperforms the best case $\nld$ algorithm in terms of speed of convergence. However, the improved asymptotic covariance matrix in the former is achieved at the cost of increased peak and average transmit power compared to the latter.

\section{Simulations} \label{sec:simulations_nld}
In this section, we corroborate our analytical findings through various simulations. In all the simulations presented, the initial samples $x_i(0) \in \R, i=1,2, \ldots, N,$ were generated randomly using Gaussian distribution with a standard deviation equal to 10. The desired global average value is indicated in each of the simulations. 
We focus here on bounded transmission functions to study their performance. Please note that our results are valid for a broader class of increasing functions (see Section \ref{subsec:nld_with_noise}) than the ones considered in this section.

\subsection{Performance of $\nld$ Algorithm Without Channel Noise}\label{subsec:perf_nld_no_noise}
\begin{figure}[tb]
\begin{minipage}{1\textwidth}
\centering
\begin{center}
\includegraphics[height=9.5cm,width=12cm]{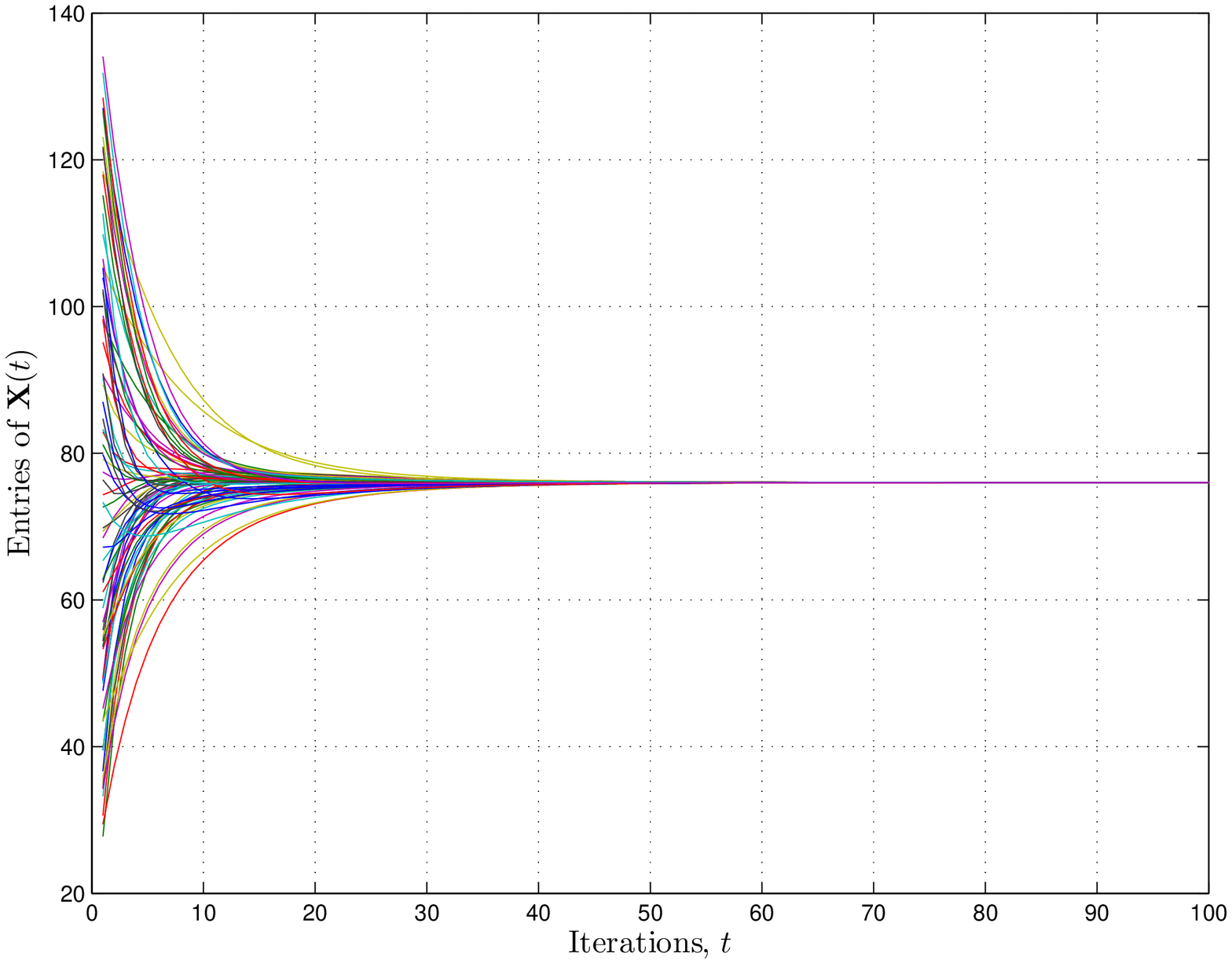}
\caption{Entries of $\mathbf{X}(t)$ versus Iterations $t$: $\alpha=1.5$, $\omega=0.01$, $N=75$, $h(x)=\tanh(\omega x)$, $\xbar=76$.}\label{fig:Consensus_LargeWSN_Noiseless}
\end{center}
\end{minipage}
\end{figure}

\begin{figure}[tb]
\begin{minipage}{1\textwidth}
\centering
\begin{center}
\includegraphics[height=9.5cm,width=12cm]{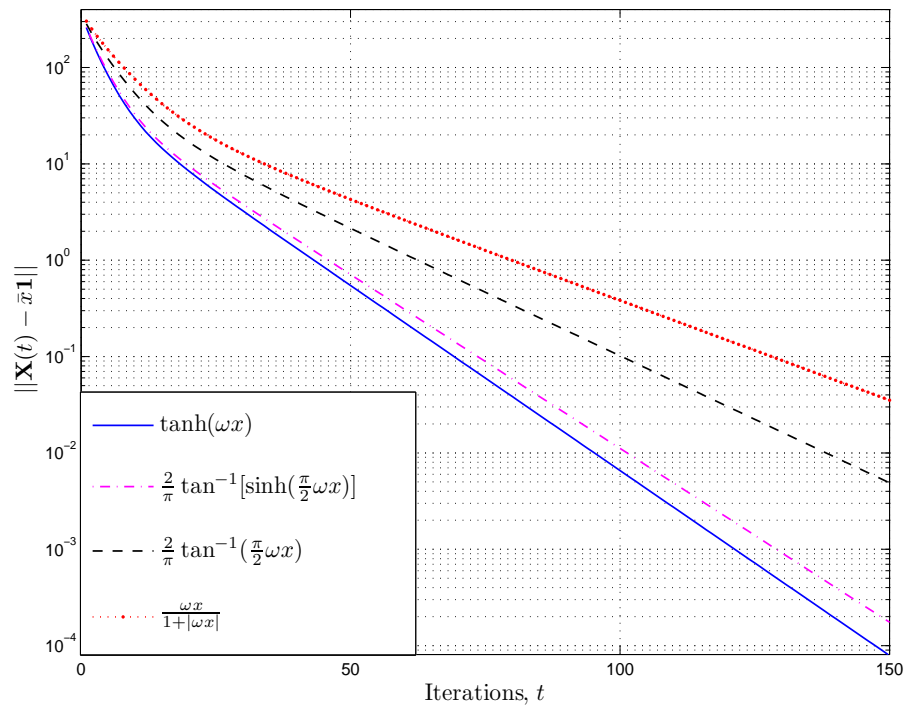}
\caption{Evolution of error $||\mathbf{X}(t)-\bar x \bf{1}||$ versus Iterations $t$: $\alpha=1.5$, $\omega=0.01$, $N=75$, $\xbar=76$.}\label{fig:ErrorNorm_LargeWSN_Noiseless}
\end{center}
\end{minipage}
\end{figure}

\begin{figure}[tb]
\begin{minipage}{1\textwidth}
\centering
\begin{center}
\includegraphics[height=9.5cm,width=12cm]{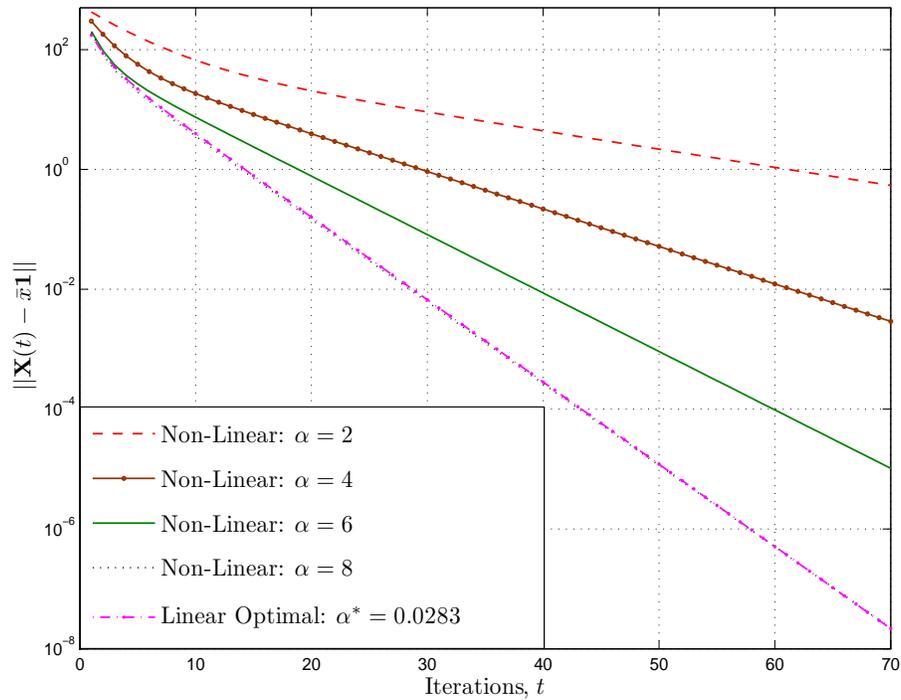}
\caption{Evolution of error $||\mathbf{X}(t)-\bar x \bf{1}||$ versus Iterations $t$: $\alpha=2, 4, 6, 8$, $\omega=0.005$, $N=75$, $h(x)=\frac{2}{\pi} \tan^{-1}[\sinh(\frac{\pi}{2}\omega x)]$, $\xbar=114$.}\label{fig:ErrorNorm_LargeWSN_alpha_Noiseless}
\end{center}
\end{minipage}
\end{figure}

Our focus in this paper is on non-linear transmissions in the presence of noise. However, we would also like to illustrate the convergence behavior on the absence of noise. Figures \ref{fig:Consensus_LargeWSN_Noiseless}, \ref{fig:ErrorNorm_LargeWSN_Noiseless} and \ref{fig:ErrorNorm_LargeWSN_alpha_Noiseless} depict the performance of the proposed $\nld$ algorithm in the absence of channel noise for a large network with $N=75$. In all the cases, we have used $\alpha$ values such that $0 < \alpha < 2 /(c \lambda_N(\La))$ as mentioned in Section \ref{subsec:mse_nld}. 

From Figure \ref{fig:Consensus_LargeWSN_Noiseless}, we infer that in about $50$ iterations, all the nodes reach consensus on the desired global average of $\xbar=76$. Figure \ref{fig:ErrorNorm_LargeWSN_Noiseless} shows evolution of error norm $||\mathbf{X}(t)-\bar x \bf{1}||$ for various bounded functions. We see that the convergence is exponential in all cases as noted in Section \ref{subsec:mse_nld}. Figure \ref{fig:ErrorNorm_LargeWSN_alpha_Noiseless} illustrates the performance of the $\nld$ algorithm when $\alpha$ is varied. Interestingly, by adjusting the step size $\alpha$ it is indeed possible to achieve the same convergence speed using the $\nld$ algorithm as that of optimal linear consensus algorithm using the Laplacian heuristic \cite{Boyd2003}. 

\subsection{Performance of $\nld$ algorithm with Channel Noise}\label{subsec:perf_nld_withs_noise}

\begin{figure}[tb]
\begin{minipage}{1\textwidth}
\centering
\begin{center}
\includegraphics[height=9.5cm,width=12cm]{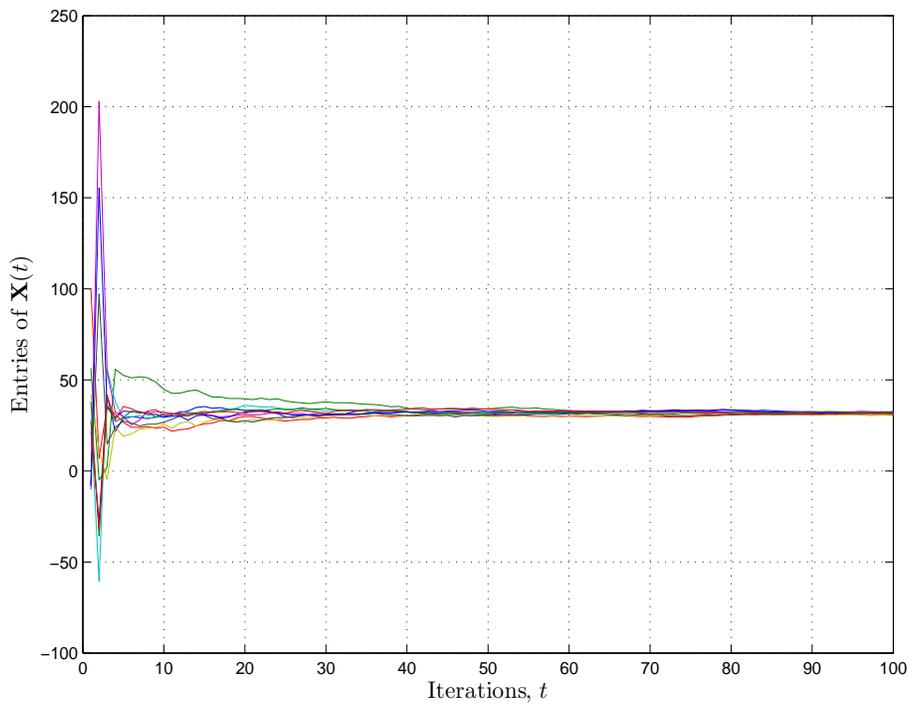} 
\caption{Entries of $\mathbf{X}(t)$ versus Iterations $t$: $h(x)=\sqrt{\rho} \tanh(\omega x)$, $\omega=0.05$, $N=10$, $\xbar=36.24$, $\rho=10$ dB, $\sigma^2_{v}=1$.}\label{fig:Consensus_SmallWSN_Noisy}
\end{center}
\end{minipage}
\end{figure}

\begin{figure}[tb]
\begin{minipage}{1\textwidth}
\centering
\begin{center}
\includegraphics[height=9.25cm,width=12cm]{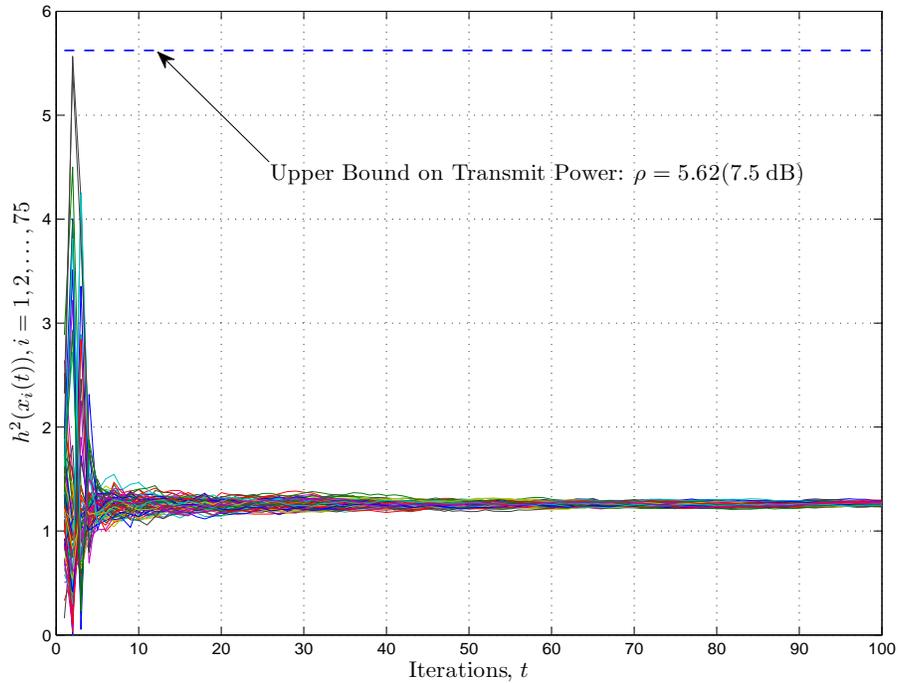} 
\caption{ Transmit power $h^2(x_i(t))$ per-neighbour versus Iterations $t$: $h(x)=\sqrt{\rho} \tanh(\omega x)$, $\omega=0.005$, $N=75$, $\xbar=102$, $\rho=7.5$ dB, $\sigma^2_{v}=0.1$.}\label{fig:LargeWSN_Transmit_Power_PerSensor_PerNeighbour}
\end{center}
\end{minipage}
\end{figure}

\begin{figure}[tb]
\begin{minipage}{1\textwidth}
\centering
\begin{center}
\includegraphics[height=9.25cm,width=12cm]{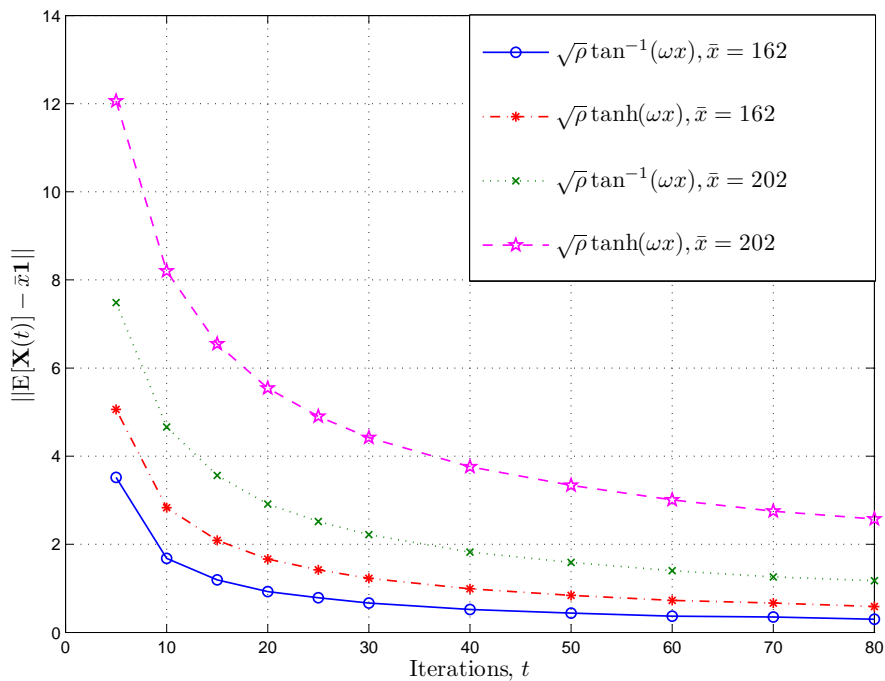} 
\caption{ $||{\rm E} [\mathbf{X}(t)]-\bar x \bf{1}||$ versus Iterations $t$: $h_1(x)=\sqrt{\rho}  \tan^{-1}(\omega x)$, $h_2(x)=\sqrt{\rho} \tanh(\omega x)$, $\omega=0.005$, $N=75$, $\xbar=162, 202$, $\rho=7.5$ dB, $\sigma^2_{v}=1$.}\label{fig:Consensus_LargeWSN_ConvMean_AsymNorm2}
\end{center}
\end{minipage}
\end{figure}

\begin{figure}[tb]
\begin{minipage}{1\textwidth}
\centering
\begin{center}
\includegraphics[height=9.5cm,width=12cm]{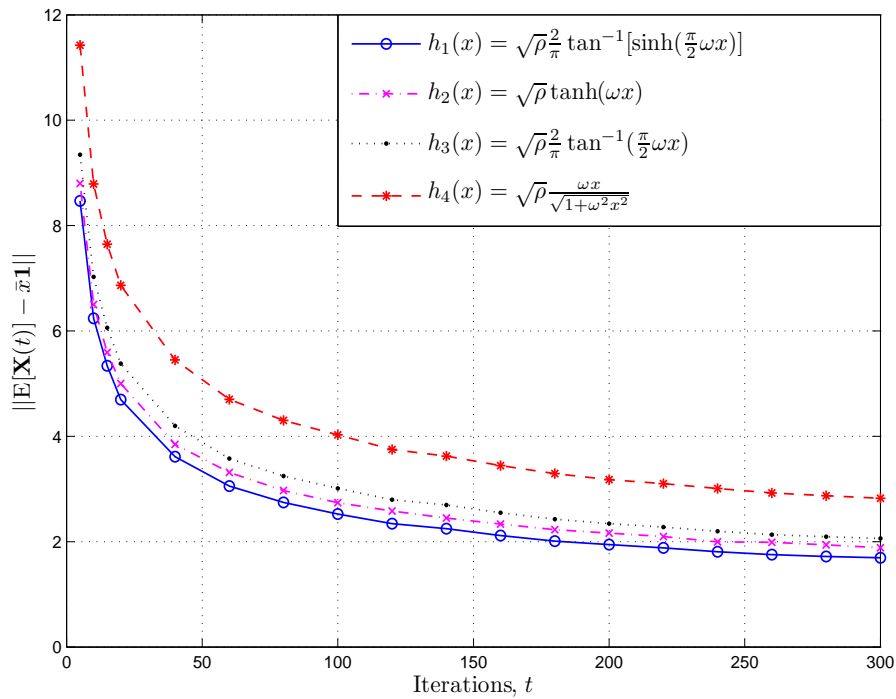} 
\caption{$||{\rm E} [\mathbf{X}(t)]-\bar x \bf{1}||$ versus Iterations $t$: $\omega=0.04$, $N=10$, $\xbar=36.24$, $\rho=5$ dB, $\sigma^2_{v}=1$.}\label{fig:Consensus_Noisy_SmallWSN_ConvMean}
\end{center}
\end{minipage}
\end{figure}

\begin{figure}[tb]
\begin{minipage}{1\textwidth}
\centering
\begin{center}
\includegraphics[height=9.25cm,width=12cm]{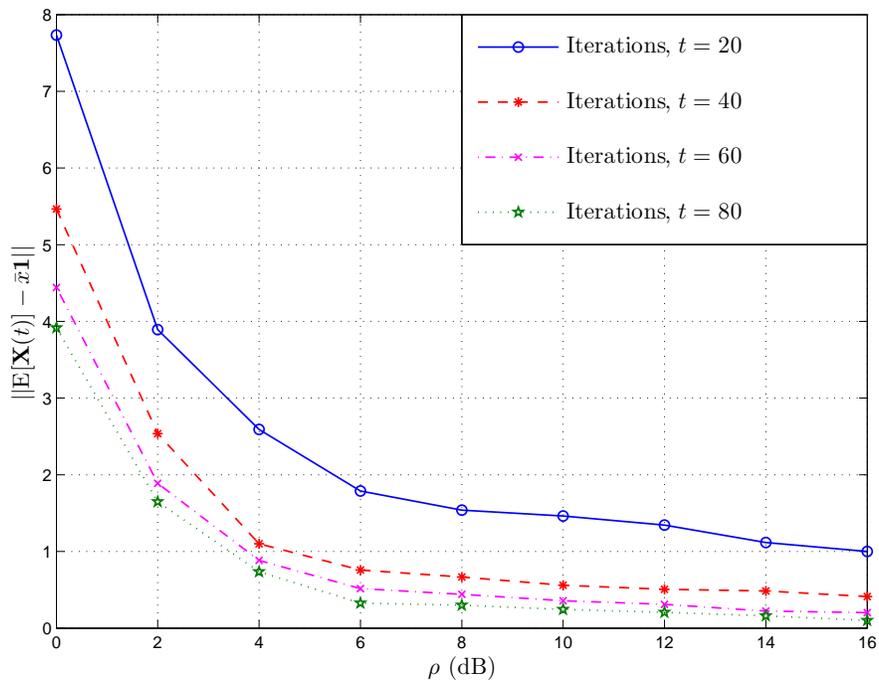} 
\caption{ $||{\rm E} [\mathbf{X}(t)]-\bar x \bf{1}||$ versus $\rho$: $h(x)=\sqrt{\rho} \frac{\omega x} {\sqrt{1+ \omega^2 x^2}} $, $\omega=0.006$, $N=75$, $\xbar=77$, Iterations $t=20, 40, 60, 80$, $\sigma^2_{v}=1$.}\label{fig:LargeWSN_ConvMean_rho}
\end{center}
\end{minipage}
\end{figure}

Figures \ref{fig:Consensus_SmallWSN_Noisy} - \ref{fig:LargeWSN_ConvMean_rho} illustrate the performance of $\nld$ algorithm in the presence of communication noise. As explained in the assumption \textbf{(A4)} in Section \ref{subsec:nld_with_noise}, we chose the decreasing step sequence to be $\alpha(t)=1/(t+1), t \geq 0$, in all simulations. Here we assumed that $\rho=\max_{x} h^{2}(x)$ is the maximum power available at each sensor to transmit its state value. Figure \ref{fig:Consensus_SmallWSN_Noisy} shows that the nodes employing the $\nld$ algorithm reach consensus for a small network with $N=10$. Figure \ref{fig:LargeWSN_Transmit_Power_PerSensor_PerNeighbour} shows the transmit power $h^2(x_i(t)), i=1,2, \ldots, N,$ per-neighbour versus iterations for a large network. Clearly, the transmit power is always constrained within the upper bound of $\rho$ (indicated by the dashed line) making the proposed scheme practically viable for the power constrained WSNs.

In Figures \ref{fig:Consensus_LargeWSN_ConvMean_AsymNorm2}, \ref{fig:Consensus_Noisy_SmallWSN_ConvMean} and \ref{fig:LargeWSN_ConvMean_rho}, we show the convergence speed performance of the proposed $\nld$ algorithm by plotting $||{\rm E} [\mathbf{X}(t)]-\bar x \bf{1}||$ versus iterations $t$. These plots indicate how fast the mean of the process $\mathbf{X}(t)$ converges towards the desired global mean vector $\xbar \onevect$. 

In Theorem \ref{asym_norm_nld_lemma}, we saw that if two functions $h_1(x)$ and $h_2(x)$ such that $h^{'}_{1}(x) > h^{'}_{2}(x), \forall x \in \R$, are employed in the $\nld$ algorithm then the convergence will be faster for $h_1(x)$ compared to that of $h_2(x)$. This is illustrated in Figure \ref{fig:Consensus_LargeWSN_ConvMean_AsymNorm2} where we have chosen $h_1(x)=\sqrt{\rho}  \tan^{-1}(\omega x)$ and $h_2(x)=\sqrt{\rho} \tanh(\omega x)$. The performance gain of $h_1(x)$ obtained over $h_2(x)$ can be understood intuitively by observing that on an average the transmit power will be more when $h_1(x)$ is employed than when $h_1(x)$ is employed. The speed of convergence for various transmit functions appropriately normalized to have the same peak power $\rho$ is shown in Figure \ref{fig:Consensus_Noisy_SmallWSN_ConvMean}. Here, we see that the transmit function $h_1(x)$ has the best performance and $h_4(x)$ has the worst performance. Intuitively this is due to the fact that $h^{'}_{1}(\xbar) > h^{'}_{2}(\xbar) > h^{'}_{3}(\xbar) > h^{'}_{4}(\xbar)$. 
Finally, we depict the convergence speed versus the power scaling constant $\rho$, the upper bound on the transmit power, in Figure \ref{fig:LargeWSN_ConvMean_rho}. For a given transmit function, increased power leads to faster convergence as would be expected, and we also observe that when the consensus iterations were increased, speed of convergence improves.

\section{Conclusions} \label{Sec:Conclusions:consensus}
A distributed consensus algorithm in which every sensor maps its state value through a bounded function before transmission to constrain the transmit power is proposed. The transmitted signal power at every node in every iteration is always bounded irrespective of the state value or the communication noise, which is a desirable feature for low-power sensors with limited peak power capabilities. In the presence of communication noise, it is proved using the theory of Markov processes that the sensors reach consensus asymptotically on a finite random variable whose expectation contains the desired sample average of the initial sensor measurements, and whose mean-squared error is bounded. The asymptotic convergence speed of the proposed algorithm is characterized by deriving the asymptotic covariance matrix using results from stochastic approximation theory. While the proposed $\nld$ algorithm has the desirable feature of bounded transmit power, it is shown that using the best case $\nld$ algorithm results in larger asymptotic covariance compared to the best case linear consensus algorithm. In the absence of communication noise, it is illustrated that the network achieves consensus on the global sample average exponentially fast provided the step size is chosen appropriately and that by adjusting the step size, it is possible to achieve the same speed of convergence as that of the best case linear consensus algorithm using Laplacian heuristic. 

\bibliographystyle{IEEEtran}
\bibliography{consensus}

\newpage

\end{document}